\documentstyle[11pt] {article}

 \title{A Variational Procedure for Time-Dependent Processes}
 \author{R. Englman$^{a,b}$ and  A. Yahalom$^b$ \\
 $^a$ Department of Physics and Applied Mathematics,\\
 Soreq NRC,Yavne 81800,Israel\\
 $^b$ College of Judea and Samaria, Ariel 44284, Israel\\
 e-mail: englman@vms.huji.ac.il; asya@yosh.ac.il;}

\begin{document}
 \maketitle

 \newcommand{\beq} {\begin{equation}}
 \newcommand{\enq} {\end{equation}}
 \newcommand{\ber} {\begin {eqnarray}}
 \newcommand{\enr} {\end {eqnarray}}
 \newcommand{\eq} {equation}
 \newcommand{\eqs} {equations }
 \newcommand{\mn}  {{\mu \nu}}
 \newcommand{\sn}  {{\sigma \nu}}
 \newcommand{\rhm}  {{\rho \mu}}
 \newcommand {\SE} {Schr\"{o}dinger equation}
 \newcommand{\sr}  {{\sigma \rho}}
 \newcommand{\bh}  {{\bar h}}
 \newcommand {\er}[1] {equation (\ref{#1}) }
 \newcommand{\gb} {{\bf \gamma}}
 \newcommand{\gcrb}  {{\bf \gamma^+}}
 \newcommand{\gd} {{\dot \gamma}}
 \newcommand{\gcr} {{\gamma^+}}
 \newcommand{\gcrd} {{ \dot \gamma^+}}
 \newcommand{\ro} {{ \gamma  \gamma^+}}
 \newcommand{\mbf} {}

 \begin {abstract}
 A simple variational Lagrangian is proposed for the time development of an
 arbitrary density matrix, employing the "factorization" of the density. Only the
 "kinetic energy"  appears in the Lagrangian. The formalism applies to pure and
 mixed state cases, the Navier-Stokes equations of hydrodynamics,
 transport theory, etc. It recaptures the Least
  Dissipation Function condition of Rayleigh-Onsager {\bf and in practical applications is
  flexible}. The variational proposal is tested on
  a two level system interacting that is subject, in one instance, to an interaction
  with a single oscillator and, in another, that evolves in a dissipative mode.

 \end {abstract}

 \section {Introduction}
 Several basic aspects of stochastic dynamics remain controversial. (A critical "state of
  art" update  in \cite{Kampen} shows this). This situation contrasts
  with most physical theories, where the problems  that arise are in the
 application of consensually accepted principles. It can perhaps be argued,
  that  the lack of  an agreed-upon variational  formulation of
  stochastic processes is at the root of the problem. {\mbf As a remedy of this
 situation, this article suggests a new variational functional
 which is to be minimized and whose minimum is the true density
 matrix.}

 To be sure, in the past
  several principles of extrema have been proposed; these include Gauss's Least
  Constraint Principle \cite {Gibbs, Lavenda}, the "least dissipation
 function" \cite {Rayleigh,O&M,McKean}, minimum entropy production-rate for steady
 states \cite {Prigogine,Callen} (see also \cite {MJK} for its
  violation), minimal energy generation rate \cite {Feynman}, minimal scattering
  integral (\cite {Kohler} - \cite{Ziman}), least velocity error during pathway \cite {Lavenda},
  the Yasue Action for stochastic mechanics \cite {BCZ}, a
 formulation involving a potential \cite {Zambrini},
  and again, recently, maximum entropy production \cite {SGS}. To these may be added
  several variational methods applicable to classical (i.e., not quantal)
 systems, such as those appropriate for general non-linear problems \cite {Tonti},
 the "governing principle for Dissipative Processes" \cite{Gyarmati, Van} and
 a generalized  Hamiltonian principle \cite {DjukicV}. Reviews of these
 and of other methods can be found in \cite{FinlaysonS} - \cite{Ichiyanagi}.

  The present proposal for a variational procedure is based on the following new elements:
  (a) the factorization (to be discussed later in this paper)
  of the density matrix as introduced by Reznik \cite {BR} and utilized recently
  by  Gheorghiu-Svirschevski \cite {SGS}, (b) a conventional Lagrangian {\mbf similar to that used in
  Mechanics to obtain the motion of a point
  particle subject to an external force}, but
  in which the scalar potential is either absent or ignorable, (c) a vector potential
 that can be singular, without this having disastrous observational consequences
 and (d) an appropriate use of minimization procedure, with origins going back
 to Gibbs \cite {Gibbs}.
 The method covers a broad range of behaviors ("deterministic" and
 stochastic, quantal and classical, electronic transport, discrete and continuous,
  markovian and otherwise) and
 places in a new perspective certain aspects in currently employed theories
 of stochastic dynamics. Apart
  from these favorable (and {\it a priori} unexpected) features, a number of
  problems remain, which will have to be resolved by future efforts.

 A pure state sub-case, {\mbf (as opposed to a density matrix that can describe a
 statistical mixture of states)}
 is the subject of a numerically worked out example in
 section 4.1 and is equivalent to the (linear) time dependent \SE. For
  this several variational formulations are known {\mbf in the literature} (\cite{Frenkel}
 -\cite{DeumensDTO}). The inter-relation between these was investigated in \cite{
 BroeckhoveLKvL}, where they were shown to be frequently equivalent. For the pure
 state case our density matrix variational method reduces to the McLachlan
 formalism \cite{McLachlan}, {\mbf in which the variation of the function is
 carried out with respect to the time-derivative only (while the function is kept
 fixed). Moreover, we give a justification of this procedure for stochastic
  processes.}

  Another application of the variational formalism, in section 4.2, includes a non-linear,
   dissipative (non-Hermitian) mechanism and exemplifies quantum jumps.
 \section {The Variation}
 The "factorized" form of the time dependent density matrix $\rho$
  (\cite{BR},\cite{SGS}) reads in terms of the column and row vectors ${\gb}$
  and (its hermitian conjugate) $\gcrb$
 \beq
 {\bf \rho =\gb  \gcrb}
 \label{rho}
 \enq
 {\mbf The above condensed notation is not trivially simple, so we give in Appendix
 A a "tutorial" on the notation.}
 We now propose an action $ S(T)$ (expressed in arbitrary units) {\mbf this being
 the integral over time $t$ with an arbitrary time end-point $T$ of a Lagrangian $\cal L$.
 It is the form of the Lagrangian that we seek: we propose that it}
  has the "quadratic" form, as follows.
 \ber
 S(T) & = & \int_0 ^{T} dt \ {\cal L}(t)
 \nonumber \\
 & = & \int_0 ^{T} dt  \ Tr \ (i\gcrd - {\bf A}^+) \cdot (-i\gd - {\bf A})
 \label {action1}
 \enr
 The variational equations based on the above action ({\mbf these are} the equations of motions for ${\gb}$ and $\gcrb$)
 are given below in \er {gd}.  Tr is the trace over all "components" of ${\bf \gamma}$ {\mbf explained in
 Appendix A}. Dots {\mbf above symbols} represent differentiation with time.

 The variational equations are obtained in the usual way by varying the action
 $S(T)$ with respect to all components of the two vectors $\gb$ and $\gcrb$.
  Thus
 \beq
 \\{\delta_{\gb}} S(T) = \int_0 ^{T} dt \delta \gb(t)[\frac{\delta{\cal L}(t)}
 {\delta \gb} - \frac{\partial}{\partial t} \frac{\delta {\cal L}(t)}
 {\delta \dot\gb}] +\frac{\delta {\cal L}(T)}{\delta \dot\gb}{\delta \gb}(T)
 \label {vari1}
 \enq
 {\mbf and a similar expression for ${\delta_{\gcrb}} S(T)$}.
 The last term, outside the integral, is the boundary term. It is assumed that
  the vectors $\gb$ are fixed initially at $t=0$, but not at any time later.
  (This will be discussed shortly.) If the boundary term can be made to vanish,
  then so will be the whole
 variation, since in the absence of a scalar potential, the
 variations $\frac{\delta {\cal L}(t)}{\delta \gb}$ and
 $\frac{\delta {\cal L}(t)} {\delta \gcr}$ vanish.  This follows,
 since for the postulated form of the Lagrangian these variations
 contain as a factor  one or the other of the
  expressions $\frac{\delta {\cal L}(t)}{\delta\dot\gb}$ and $\frac{\delta
  {\cal L}(t)}{\delta \gcrd}$ and these vanish due to the boundary
  variation. {\mbf(In fact, $\frac{\delta {\cal L}(t)}{\delta\gb}$ and $\frac{\delta
  {\cal L}(t)}{\delta \gcr}$ also vanish due to the presence of the same
  factors, but these variations do not form a sufficiently general
  basis for the variation procedure, since the vector potential
  $\bf A$ may not be a function of {\it all} components of the $\gamma$-vector.)}

 At this point the role of the boundary term is well worth reflecting upon.
 {\mbf It is not present in, e.g., deterministic Mechanics, where the values
  of the variables are fixed at both end points.}
 However, it is well known that
 the boundary term arises when the value of the variant quantity, i.e.
 $\gb$, is undetermined at a boundary.  This is (physically) the case when
  a random force operates
 on the system. Thus, we are not allowed to neglect the boundary term. It is
 now a further "bonus" in the formalism, that the {\mbf vanishing of the
 boundary-term variation does not interfere with the vanishing of the body
  terms-variation, (i.e., it is neither contradictory to it nor incomplete
 with respect
  to it), but is by virtue of our choice of the Lagrangian
  precisely} identical with it.

 In summary, we have the variational equations
 \beq
 \frac{\delta {\cal L}(t)}{\delta \dot \gb}=0
 \label{var2}
 \enq
 and their complex conjugates
 \beq
 \frac{\delta {\cal L}(t)}{\delta \gcrd }=0
 \label{var2cc}
 \enq
 and these make the action extremal (and an absolute minimum) also when
  the "vector potential"\footnote{We have named our frequently used quantity
   ${\bf A}$ the "vector potential", in analogy
with the quantity that enters as a cross-term with the time
derivative ("the particle velocity") in the Lagrangean of
classical mechanics \cite{LandauL}, or inside the square with the
canonical momentum and in distinction from the scalar potential
$\phi$, which appears in (non-relativistic) Lagrangeans as a
separate term.} $\bf A$ is a function of $\gb$ \cite{LandauL}.[To
see this,
  note the remark after \er{gd}]
  This has the immediate consequence that
  in the expansion of the Lagrangian, shown in \er{action1}, the following
 expression needs to be minimized:
 \beq
 'DF'=\gcrd \gd -i[\gcrd {\bf A} - \gd {\bf A^+} ]
 \label{DF}
 \enq
 This follows, since the ${\bf A \cdot A^+}$ term is independent
  of time-derivatives and is not varied. We further note that (by the form
  of the vector
  potential) $ i $ times the square bracket is a real quantity. The quantity in
 the above equation is essentially of the form of  Onsager and
 Machlup's Dissipation Function
 [the negative of eq. (4-25) in \cite {O&M}. One recalls that their
  Dissipation Function is also minimized only with respect to the time
 derivatives of the variables {\mbf just as in the procedure proposed
 in \cite{McLachlan} and in the present work}]. To bring  our 'DF' precisely to the form
 of the Dissipation Function, we need to go from our variables $\gb$ and $\gcrb$ by
 a constant linear transformation (not necessarily a unitary one) to the variables
 ${\bf \alpha}$
 of \cite {O&M}. One will then get, instead of the diagonal form $ \gcrd \gd $,
  a non-diagonal form which defines the "generalized resistance matrix"
 $R_{ij}$ of \cite {Lavenda, McKean}. [Since the use of extensive variables is to be
  preferred  to intensive ones (and $\gb$ and $\gcrb$ are of the latter type)
  the  transformation should include the system size. Alternatively,
 the action integral may be premultiplied by a size-dependent scale factor.]

 Thus, the result of the variation are a set of equations:
 \ber
 \gd & = & i{\bf A}
 \nonumber \\
 \gcrd & = &-i {\bf A^{+}}
 \label {gd}
 \enr
 When $\bf A$ and $\bf A^{+}$ are functions of $\gamma$ and $\gcr$, their (non-zero)
 derivatives come in with either $\gd - i{\bf A}$ or $\gcrd +i {\bf A^{+}}$ as factors and
 these factors vanish due to the above equations.

 [We can illustrate this in the case of two components, for
which we write the Lagrangian in equation (1) as \beq {\cal L}(t)
=f_1^{+}f_1+f_2^{+}f_2\label{Lag2}\enq and $ f_1=-i{\dot\gamma}_1
- {\bf A}_1(\gamma_k,\gamma_k ^{+})$ etc.  with the $\gamma$
dependence explicitly put in the $A$'s. Recall that equation (7)
means that for the variational solution
$f_1=f_1^{+}=f_2=f_2^{+}=0$ at all times and that with this
solution the minimal action is zero.

Upon varying with respect to e.g., $\gamma_1$, the resulting
Lagrange-Euler equations is now in full detail \beq
-(\partial_{\gamma_1}A_1^{+})f_1 +i{\dot
f}_1^{+}-f_1^{+}(\partial_{\gamma_1}A_1)-(\partial_{\gamma_1}A_2^{+})f_2
-f_2^{+}(\partial_{\gamma_1}A_2)=0\label{LE}\enq Since all the
terms contain one of the $f$-factors or their time derivatives
(which are necessarily also zero) the above equation is satisfied
for the proposed variational solution. There may be other
solutions, though! If for these not all $f$-s are zero (and
therefore differ from the proposed solution), than the action
(which consists of positive terms) is positive and larger than
that for the solution given in \er{gd}.]

  These are the equations of motion of the (independent) vector variables
 and can be regarded as having the status of the Langevin equations, or
  Hamilton's equations for the set of conjugate variables $\gamma_a $ and
  $ \gamma^+_a $.
The processes considered in this section comprise a purely
Hamiltonian process $A_H$ {\mbf (namely, energy  preserving,
 "elastic") as well as}
  some other, dissipative mechanisms. Thus, a Markovian scattering process
  represented by the symbol $M_{ba}$ (designating half the probability per
  unit time of {\mbf a scattering event taking the system from state $a$ to
  $b$)} can  be written
  as $(A_{out},A_{in})$ to separate scattering out of and into a given state.
  We also add a stochastic,
 random process arising from, e.g., an external source, as $A_r$. {\mbf Some
  other type of processes will be considered below in section 5.}

  For the  first two processes we have
  \ber
  A_H & = & -H\gamma
\nonumber\\
  A_H^{+} & = & -\gamma^{+} H^{+} = -\gamma^{+}H
  \nonumber\\
  (A_{out})_a & = & i\sum_{b}M_{ba}\gamma_{a}
 \nonumber\\
  (A_{out}^{+})_a & = & -i\gamma_a^{+} \sum_{b}M_{ba}
  \nonumber\\
  (A_{in})_a & = & -i\frac{1}{N}(\sum_b M_{ab}\gamma_b
  \gamma^{+}_b)(\gamma_a^{+})^{-1}
 \nonumber\\
 (A_{in}^{+})_a & = & i(\gamma_a)^{-1}(\frac{1}{N}\sum_b M_{ab}\gamma_b \gamma^{+}_b)
\label{A}
\enr
When we add to these the random force, we obtain in
addition
\beq
(A_r)_a =-if_a\gamma_a \qquad
(A_r^{+})_a=i\gamma_a f_a
\label{random}
\enq
$f_a$ represent the components of the random time-dependent force
with zero mean and a finite self-correlation. ($N$ is the number of
states in the ensemble, see appendix A)

  The vector potential $\bf A_{in}$ is singular.
 However, singularities in vector potentials are well known (as,
 e.g., in those for solenoidal or monopole fields).
  To cancel these singularities
 we shall follow the procedure of Reznik \cite {BR} and Gheorghiu-Svirshevski
  \cite {SGS}, who multiply $\gcrb$ into
 $\gd$, $\gb$ into $ \gcrd $, add and obtain the (master) equations
 for the density matrix.

 After substituting the quantities from \eqs (\ref{gd},\ref{A},\ref{random})
 we obtain {\mbf by this procedure} for a diagonal
 element  (say) $aa$ of the density matrix
 \beq
 \dot\rho_{aa}  =  (\gd \gcrb +\gb \gcrd)_{aa}
  =  i([\rho,H])_{aa}  - 2 \rho_{aa} \sum_b M_{ba}+ 2 \sum_b M_{ab}\rho_{bb}
  \label {diagonal}+2\rho_{aa} f_a
 \enq

 {\mbf When one writes out the equation, similar to \er{diagonal}, for the
  time derivatives of the off-diagonal matrix elements $\rho_{ab} (a \neq b)$,
  one finds singularities in them, due to the above mentioned singularities in the vector
  potentials.
  For a macroscopic system these singularities cancel,
  when one takes into account the phase decoherence between different states of
  macroscopic bodies.} (The
 subject of microscopic to macroscopic transitions does not belong here. It
  was studied by various methods and has summaries in e.g. \cite {Balian,
  Chester}.)
 \section {Potential Fluid Dynamics}
 An interesting application of the preceding complex factor-density formalism
 is for the well known potential flow  (namely, fluid dynamics without vorticity) as presented in
 many hydrodynamic text books, e.g., \cite {Lamb}. If a flow satisfies the condition of
  zero vorticity, i.e. the velocity field $\vec v$ is such that
 $\vec \nabla \times \vec v = 0$, then there exists a function $\phi$ satisfying
 $\vec v = \vec \nabla \phi$.

 {\mbf In this section we answer the question: what form of} the vector
 potential ${\bf A}$ appearing in \er{action1} will ensure, that upon variation
 {\mbf of the action containing these vector potentials} we  shall
  obtain precisely {\mbf the well known equations of potential flow hydrodynamics. These
 equations are:}
 \ber
 \frac {\partial \rho }{\partial t}  & + &
 \vec \nabla \cdot (\rho \vec \nabla \phi)  =  0
 \label{ContinS2p}
  \\
 \frac {\partial \phi}{\partial t} & = & -\frac{1}{2} (\vec \nabla \phi)^2 - h -
 \Phi -\nu \nabla^2 \phi
 \label{nueq2p}
 \enr
 In these equations {\mbf the physical meaning of the quantities is that }
 $\rho$ is the mass density, $h $  is the specific enthalpy,
 $\nu$ is the viscosity coefficient
  and $\Phi$ is some function representing the potential of an external force acting
  on the fluid. The first of these \eqs is the continuity
  \eq, while the second is a modified Bernoulli's equation which takes
  into account some viscous effects (A full viscous flow is of course not a potential
  flow and contains vorticity) . $\rho$ and $\phi $ play the
  roles of the squared-amplitude and of the phase angle, respectively. Both are real
 quantities.

 {\mbf The final results for the desired vector potentials and their complex conjugates
 are shown, below, in \er{gdot2} and \er{gdot3}. To obtain them, we first express
 the variational variables $ \gamma $ and $ \gamma^+$ that we have used so far
  in terms of the physical variables $(\rho,\phi)$. The variation will now be
 carried out with respect to the latter variables. The transformation is}
 \ber
 \gamma &=&\sqrt\rho e^{i\phi}
 \label{gamhy1}
  \\
 \gamma^+&=&\sqrt\rho e^{-i\phi}
 \label {gamhy2}
 \enr
 Though all variables are now functions of the positions, and are thus continuous
  variables, we shall label them, as before, by the subscripts $a$, etc. The
 following relations  (with no summations over repeated symbols) {\mbf
  arise simply from the inverse transformation}:
 \ber
  \dot\phi_a & =& (\gd_a \gcr_a -\gamma_a \gcrd_a )/2i\rho_{aa}
 \label{nud}
 \\
   \vec\nabla \phi_a & = & (\vec\nabla \gamma_a \gcr_a -\gamma_a \vec\nabla\gcr_a )/2i\rho_{aa}
 \label {nudelta}
 \\
 \nabla^2 \phi_a & =& (\nabla^2 \gamma_a \gcr_a -\gamma_a \nabla^2\gcr_a )/2i\rho_{aa}
 \nonumber\\
 & &-((\vec\nabla \gamma_a \gcr_a)^2 -(\gamma_a \vec\nabla\gcr_a)^2 )/2i\rho_{aa}^2
 \\
 \vec \nabla \rho_{aa} & = &(\vec\nabla \gamma_a \gcr_a + \gamma_a \vec\nabla\gcr_a )
 \label {rhodelta}
 \enr

 Using these, we first rewrite \er {ContinS2p}  as
 \ber
 \dot \rho_{aa} & = & (\gd_a \gcr_a + \gamma_a \gcrd_a)
 \label {C1}
                 =  -\rho_{aa} \nabla^2 \phi_a -\vec\nabla \rho_{aa} \cdot  \vec\nabla \phi_a
 \label {C2}
 \nonumber\\
   & \equiv & R_a (\gamma_a , \gcr_a )
 \label {ContinS3p}
 \enr
 $R_a $ being a well-defined, real function of the variational variables
  and of their first and second spatial derivatives (but independent of
  the time-derivatives). Likewise, one obtains as the rate equation for
  the phase, \er {nueq2p}, as
 \ber
 \dot \phi_a & = & - (\gd_a \gcr_a -\gamma_a \gcrd_a)/2i\rho_{aa}
 \label {Cnu1}
 \nonumber\\
   & = & -\frac{1}{2} (\vec \nabla \phi)^2 - h -\Phi -\nu \nabla^2 \phi
 \label {Cnu2}
 \nonumber\\
   & \equiv &  N_a (\gamma_a , \gcr_a )
 \label {nueq3p }
 \enr
 in which $N_a $ has properties similar to $R_a $.

 We can solve  for the two quantities, $\gd_a \gcr_a $ and $\gamma_a \gcrd_a $
  from the preceding  two equations, and then divide by  $\gcr_a  $ and
  $\gamma_a $,  respectively, to obtain the time derivatives. However, by \er{action1}
 the time derivatives are just the vector potentials. Thus we finally obtain
 \ber
 \gd_a  =  (R_a +2i\rho_{aa} N_a )\frac{1}{2\gcr_a }
        =  iA_a
 \label {gdot2}
 \enr
 and the complex conjugates
 \ber
 \gcrd_a   =  \frac{1}{2\gamma_a }(R_a -2i\rho_{aa} N_a )
        = - iA_a^+
 \label {gdot3}
 \enr
 {\mbf We have thus found the vector potentials which have to be inserted in
 the action, so as to yield variationally the hydrodynamic equations,
 \er{ContinS2p} and \er{nueq2p}.}
 It is evident that the complex representation is a natural way to obtain
 variationally equations of motion for two such dissimilar quantities as amplitude
 and velocity. {\mbf The physical extrema are certainly global
  minima (although the functional may have additional minima).
  Detailed applications will be undertaken in the future including the problem of
  a general viscous flow}.

\section {Applications}
\subsection {A Periodically Varying Hamiltonian}
 We shall now apply the proposed variational procedure to yield, {\mbf in
  one case} exactly {\mbf and in another case}
  approximately, the solution for a (hermitian) Hamiltonian that
 has a periodic variation. The cases  chosen are  such that analytical
  solutions are
 known exactly (\cite{EYB1998}-\cite{EYACP}), so that we can compare to them
  the variational solutions to be obtained here.
  Specifically, we consider the time
  development of a doublet  subject to a Schr\"odinger equation whose
  Hamiltonian  in a doublet representation is
 \beq
 H(t)= G/2\left( \begin{array}{cc}
   -cos(\omega t) &  sin(\omega t) \\
   sin(\omega t)  &  cos(\omega t)
 \end{array} \right)
 \label{hper}
 \enq
 Here $\omega $ is the angular frequency of an external disturbance.
 The eigenvalues of \er{hper} are $\frac{G}{2}$ and $-\frac{G}{2}$. If $G>0$, then in the
 ground state the amplitude in  the upper component {\tiny $ \left(\begin{array}{c}
  1 \\ 0  \end{array} \right) $}  in  \er{hper} is
 \ber
  C^u_g   &=&   \cos (Kt) \cos ( \omega t/2) + (\omega/2K) \sin(
 K t) \sin (\omega t/2)
 \nonumber\\
         &+& i(G/2K) \sin(K t) \cos( \omega t/2)
 \label{CGspec}
 \enr
 with \beq K =0.5 \sqrt{G^2 + \omega^2}
 \label{K} \enq The amplitude $C^l_g$ of lower component {\tiny $
 \left(\begin{array}{c}
  0 \\ 1  \end{array} \right) $} in the ground state has a similar form, which
  we shall not bother to write out. For the variational procedure we postulate
 a superposition of complex circular functions
 \beq
 C^u_g = \sum_{m=-M,M+1}A^u_m exp^{i(m-\frac{1}{2})\omega t}
 \label {superposition}
 \enq
 and similarly for $C^l_g$. The complex coefficients $A^u_m$ and $A^l_m$ are
  determined by minimization
  of the variational action, \er{action1}, subject to the normalization condition
 that the sum of the absolute squares of the coefficients is unity.  $m$ takes
  integral values between the limits and we have taken for our trial functions
  $M=2$, that is six terms in each component. The half integer
 in the exponent is suggested by the acquisition of a Berry phase after a full
  period $\frac {2\pi}{\omega}$. For the same reason we have taken the range
  of the integration in \er{action1}
 to be twice the period.  So as to create realistic conditions for the
  implementation of the variational procedure, we have
 chosen a {\it finite} range for the time variable, although, as can be seen
  from the  formulae in \er{CGspec} and \er{K}, for a general value of $G$ the
 solution is not time-periodic. We illustrate the procedure for two cases.

 \subsubsection{A periodic case}

 This comes about when $G$ is such that the Rabi frequency  $K$ in \er{K} and
 $\omega$ are commensurate. We have chosen $G=\sqrt{15}\omega$, so that $K=2\omega$.

 \begin{figure}
 \vspace{4cm}
 \begin{picture}(1,1)
 \end{picture}
 \includegraphics{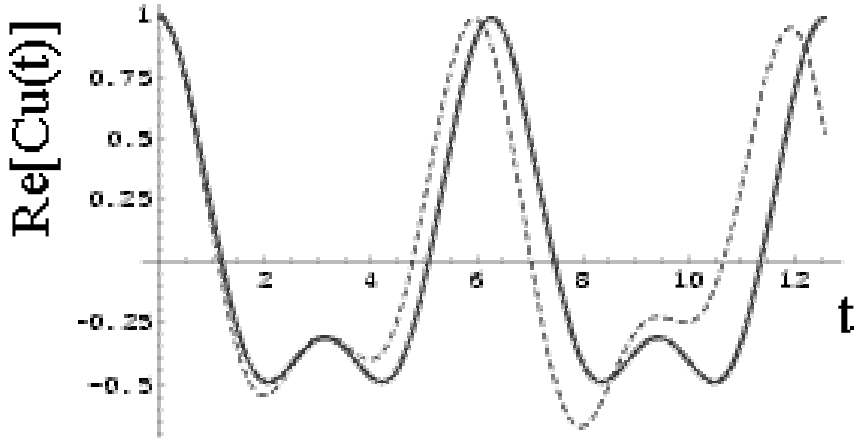}
 \caption {Time-dependence of the real part of the amplitude in the
 upper electronic component state. Parameter values $G=3, ~ \omega
 =1$. Full line: variationally obtained state. Broken line: the
 exact, algebraic solution.}
 \end{figure}
\vspace{2cm}
 \begin{figure}
 \vspace{6cm}
 \begin{picture}(1,1)
 \end{picture}
 \includegraphics{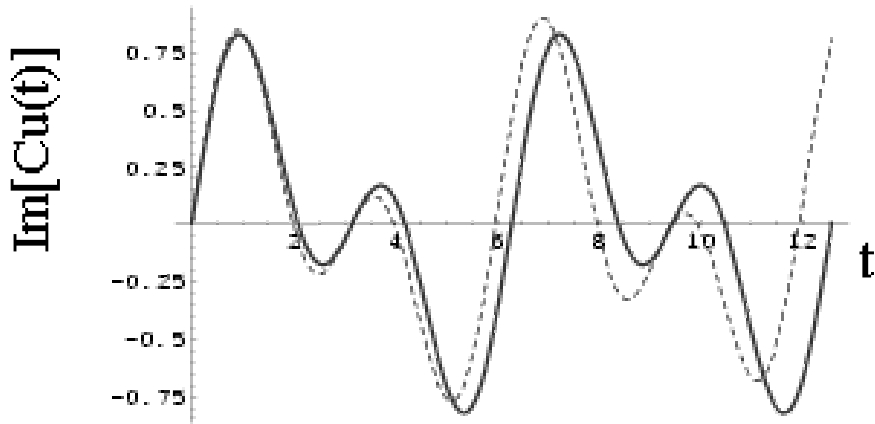}
 \caption {Same as in the previous figure but for the imaginary
 part of the amplitudes.}
 \end{figure}

 Minimizing the action with respect to the coefficients subject to
 the normalizing conditions turns out to be equivalent to
 diagonalizing a twelve-by-twelve matrix, whose elements are the
 action integral computed with the circular function shown in
 \er{superposition}. There are twelve-by-twelve matrix elements,
 rather than just six-by-six, since the upper and lower states are
 coupled by the off-diagonal terms in \er{hper}.

 We find a pair of zero eigenvalues (actually eigenvalues of about $10^{-15}$),
  whose meaning is the value of the action integral in the transformed
  representation; in other words, our variational solution is exact. Also, the
  numerical values  of the variationally obtained coefficients $A^u_g $ agree with
  those in the analytic solution in \er{CGspec}. For comparison,the other
 eigen solutions have "eigenvalues" of orders 1-20.

 \subsubsection{A non-periodic  case}

 With the choice of e.g., $G=3$ and $K=\frac{\sqrt{10}}{2}$, the analytic solution
 shown in  \er{CGspec} is not periodic and an exact solution cannot be achieved
 variationally while having a finite $t$-range in the action integral. Moreover,
 a larger spread of the basic set in \er{superposition} is needed. Still,
 so as to estimate the efficacy of the variational procedure under non-optimal
  conditions, we used the same $t$-range and the same set size as in (a).
 The lowest eigenvalues ($=$ the values of the action integral) are about $0.35$,
  compared with others eigenvalues, which are again in the range of 1-20.

 The results are also shown graphically, by comparing the variational solution
  (full lines) with the exact, analytic solution (broken lines) in \er{CGspec}
 for real and imaginary parts in Figures 1 and 2, respectively. The similarities
 are quite good for the first half (that comprises the period of the Hamiltonian),
 but is worse in the second half and further deteriorates later, {\mbf say in
 the time range $ ({\frac{4\pi}{\omega}},{\frac{6\pi}{\omega}})$. On the
  other hand, had we taken the time range of the variational procedure (the
  upper limit of integration $T$ in \er{action1}) up to ${\frac{6\pi}{\omega}}
  $, we would have obviously got a somewhat different solution, which would have
 improved the approximation in the latter range and spoiled somewhat the agreement
 in the earlier range. In general, the  approximate solution depends on the
 time-range of the action. In our view, this endows a flexibility to the
 practical application of the method,
 in the sense that, depending on which time range is of interest, better
  approximation can be achieved for that range. Of course, when the "approximate"
 solution is identical with the true solution in some range, it will remain so,
 by analytical continuation, for all ranges.}

 In conclusion,
  one notes a successful application of the variational principle for a purely
  hermitian case, whose solution, though available by algebra, is not trivial.

  \subsection{A Non-linear Evolution}

  The continuous passage of an initially prepared pure state to transitions resembling
  quantum jumps was recently studied in \cite{RauW}, based on a form of the Liouville-von Neumann-Lindblad (LvNL)
  equation. The actual form used originated in a representation of
  fast level crossing in molecular systems involving two states \cite {Kayanuma}. It was noted
   in \cite{RauW} that  the factorization formalism, called the" square root operator"
  method of \cite {BR}, represents an alternative way of showing how a dissipative term in the
  Hamiltonian can cause decoherence. To apply our variational formalism to this case, we first
   formulate the evolution equations in the factorization scheme and  solve
   the resulting equations (this is done (A) below) and, secondly, we obtain an approximation
   to the solution by minimizing the action with respect to some parameters appearing in the
   assumed $\gamma$'s (this is carried out in (B)).

 \subsubsection{Decoherence by the "square root operator" method}

The vector potentials
consisting of a Hamiltonian and a dissipative (non-Hermitian) part are now written for the two
 factors $(\gamma_1,\gamma_2)$ of the density matrix as
\ber {\bf A}_1 & = &-\frac{1}{2}G cos(\omega t) \gamma_1 -
J\gamma_2 -i\Gamma[\gamma_1
-|\gamma_2|^2(1+ \mu|\gamma_1|^2)/\gamma_1^{+}\nonumber\\
{\bf A}_2 & = & \frac{1}{2}G cos(\omega t) \gamma_2 - J\gamma_1
-i\Gamma[\gamma_2  -|\gamma_1|^2(1-
\mu|\gamma_2|^2)/\gamma_2^{+}\label{dissipa}\enr One notices the
similarity of these expressions with the corresponding formulae in
\cite{Kayanuma} and \cite {RauW} (where the interpretation of the terms is spelt out)
and also the divisor $\gamma$ on
the extreme right, characteristic of the factorization formalism
for dissipative processes, \er {A}. The trace of the density
$(\rho=\gamma\gcr)$ stays constant during the motion and this property is maintained unchanged
also by the superlinear terms which enter with the coefficient $\mu$.

It may be noted that the equations of motions for $\gamma_1(t)$ and $\gamma_2(t)$ and of their
conjugates, given in \er{gd}, lead to the master equations for the matrix elements of the density operator.
For the diagonal elements these are of the LvNL form (when $\mu=0$), but not for the non-diagonal
ones. This property  has already been noticed in \cite{BR}.

We next solve two equations for the $\gamma's$, subject to the pure state initial conditions
$\gamma_1=1$, $\gamma_2=0$ at $t=0$, then form from the solutions
the diagonal matrix elements of the density matrix and finally show the results in figure 3.
 The quantity
changed between the upper three drawings is the strength $\Gamma$ of the dissipative term.
 As this
increases, a transition takes place from the slow to the fast decoherence regime. We note the remarkable
   similarity of the results obtained here by the factorization
method to those in Figure 1 of the above articles, except that for strong
dissipation their drawings show little oscillations, unlike our third drawing from above.
 In this, drawn for $\frac{\Gamma}{\omega}=20$,
 after a very steep initial slope (not visible in the figure), both diagonal density matrix
  elements oscillate about the asymptotic value of $\frac{1}{2}$.

  The majority of calculations whose results are shown in this section are carried out for
  a density matrix referring
  to  an "assembly" consisting of one system. This means that in Appendix A, one has $N=1$ and
  the summation over $\alpha$ in \er {rhomatrix1} is trivial. We have also worked the density
  matrices when there is a non-trivial summation, namely when initial conditions on the two
  factors are $\gamma_1(0)=e^\frac{i\pi}{2}, e^\frac{2i\pi}{2}, e^\frac{3i\pi}{2},
  e^\frac{4i\pi}{2}$ and  $\gamma_2(0)=0$, (so that $N=4$), instead of having only
  $\gamma_1(0)=1$ (and $N=1$),
   as before. The resultant density tends now to an almost perfectly straight line. This is similar
   to the graphs shown in both \cite{Kayanuma} and in \cite {RauW} for strong dissipation and
   elucidates the meaning of system-averaging in the density matrix. (In more
   complex systems, one would require an averaging in the density matrix over a much larger
    number of states,  such that $N \to \infty$.) We have also worked out the $N=4$ case for
    the upper three graphs in figure 3. For these graphs, there was hardly any perceptible
    change from  those shown.

 For the detailed interpretation of these results, which is not the primary subject of this
  work, we again refer to \cite{RauW,Kayanuma} and references therein. Also, we do not delve here into the details or
 extensions of the results, but turn to the variational treatment of time-development to be
 got from minimizing the action.
\begin{figure}
 \vspace{12cm}
 \includegraphics{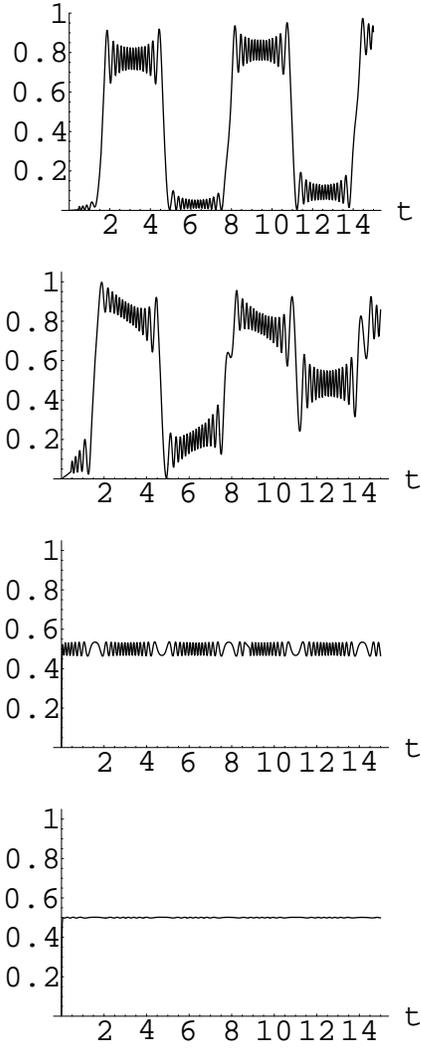}
 \caption {Unitary evolution. The diagonal element of the density matrix $\rho_{22}$ under a
 successively increasing dissipative parameter $\Gamma$. The parameters in \er{dissipa} are
 $G=45$ (25 in the bottom drawing, chosen to facilitate the computation), ~$J=3$,~$\omega=1$,
 ~ $\mu$ (superlinearity coefficient)$=0$. Then, in top drawing: $\Gamma=0$, $N$ (Ensemble size)$=1$;
  in second drawing: $\Gamma=0.05$, $N=1$; in third drawing: $\Gamma=20$, $N=1$; in bottom
   drawing  $\Gamma=20$ (as in previous, but)~ $N=4$. The initial slope in the last two drawings is too
   steep to be visible and so are tiny fluctuations in the horizontal part of the bottom drawing.
   A non-zero value of the superlinear parameter $\mu$, of the order of $1$, hardly changes
    the curves. }
 \end{figure}

 \subsubsection{Solving variationally}

We note that the strongly oscillatory factor in the solution arises from the driving
term $\frac{G}{2} cos (\omega t)$ and that this term was already present in the Hamiltonian case
considered in the previous subsection, in \er{CGspec}. (We have, however, eliminated there the fast
oscillating factor by subtracting from the Hamiltonian the so-called dynamic-phase.) So in this
subsection we shall put $G=0$, which also makes the numerical aspect of the variation considerably
simpler. We then set up pair of  suitable trial $\gamma$(t)'s, containing parameters to be varied.

 In contrast to the previous subsection (which was a linear problem and in which a large number of
  Fourier coefficients were varied), in the present problem only one variational parameter ($v$)
  is introduced. However, to make progress, we must consider critical regions of the time domain,
  namely  $t=0$ and $t=\infty$. At the former, it is easy to see that in order that the singularity
  due to the zero divisor $\gcr_2(0)$ in the vector potential be matched by the time derivative
  of $\gamma_2(t)$ at $t=0$, this function must take there the form of
  \beq
  \lim_{t\to 0}~\gamma_2(t)\to \sqrt{4e^{i\beta}\Gamma t}+O(t)\label{g20}\enq with the constant
   phase angle $\beta$ arbitrary.
Similarly, it can be shown that, asymptotically for large $t$, the same solution must have the form
\beq \lim_{t\to \infty}~\gamma _2(t) \to e^{-iJt}\sqrt{[1+ ve^{-2(\Gamma-iJ)t}+o(e^{-2\Gamma t})]/2}
\label{asympt}\enq or some other form equivalent to this. A constant phase factor was
ignored here. The parameter $v$  cannot be found from the equations of motion.
 We seek to obtain the variationally best $v$, such that the condition at $t=0$ is also satisfied.
 After some elementary algebra on finds that in terms of the variational parameter, the density
  factor $\gamma_2$ can have the form,
  \beq\gamma_2(t)=e^{-iJt}\sqrt{[1+ v e^{-2(\Gamma-iJ)t}-(1+v)e^{-((v+2)
  \Gamma-ivJ)\frac{2t}{1+v}}]/2}\label{asympt2}\enq
  The previously introduced phase $\beta$ was varied independently and found to be
  small. So we put $\beta=0$. The other density factor $\gamma_1(t)$ was so constructed
  that the exponentials inside the square root had the opposite signs to those in $\gamma_2$
  and normalizing factors were added so that $|\gamma_1|^2 + |\gamma_2|^2 = 1 $ at all times.

  Minimizing the action integral for a set of parameter values for $J$ and $\Gamma$
  yields optimized $v$'s.  Keeping $J=3$ fixed (as in references \cite {Kayanuma,RauW}
  and selecting a set of dissipation parameter we obtain  as follows:

  $J=3,~\Gamma=~0.1,~ v~=~-.985$

   $J=3,~\Gamma=~0.5,~ v~=~-.78$

   $J=3,~\Gamma=~2,~~~ v~=~-0.755$

   For the middle case $\Gamma=.5$ we compare in Figure 4. the variational solution
   (thin continuous curve, with $v=-.78$) and the
   solution from the equation of motion (thick continuous curve). In the
    asymptotic regime of large $t$, the behavior of the two curves is quite similar, though
    the amplitude of oscillation is clearly smaller in the variational solution than in the
   exact solution. The discrepancy appears the more serious in that a different choice of
  the variational parameter, namely $ v=-0.98$, (also shown in the figure by broken lines),
   which has an action larger than the optimized one, comes nearer to the exact solution.
   However, we show in the inset that in the extremely short time region, the optimized solution
   is {\it qualitatively} better than the other choice. Because of the singular
  behavior of the  short time region in the vector potential, this region dominates the value of the
  action. At the same time, yet another choice of the parameter $v=-0.05$, (shown in the figure
   by dotted lines) gives a definitely poorer resemblance to the correct curve.

  In conclusion, when it comes to describe subtle quantum mechanical ensemble
  properties, the factorization (or "square root operator") method can be used either
  in its equation of motion form or variationally. In  the present case, at least, the
  equation of motion approach was from a numerical view-point undoubtedly superior. So one may
  question the use of the variational method. However, on the one hand not all problems may
  be easily solvable. On the other hand,
  one should remember that the derivation of the equation of motion in \er{gd} is itself
   based on a variational ansatz introduced in this paper.
 \begin{figure}
 \vspace{6cm}
 \begin{picture}(1,1)
 \end{picture}
 \includegraphics{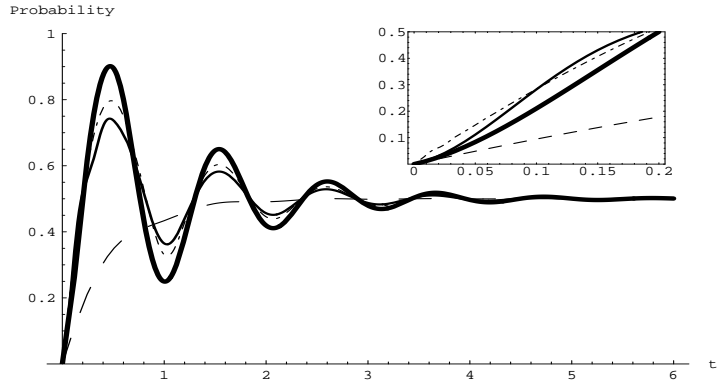}
 \caption {Comparison of solutions.
 ~ $\rho_{22}$, a diagonal component of the density matrix,
 for the following values of the parameters in equation \ref{dissipa}: $G=0$,~$J=3$,~ $\omega=1$,
 ~$\Gamma=0.5$,~$\mu=0$. Solution of equation of motion: thick line.
 Variational solution in \er{asympt2} with optimized  parameter $v=-0.78$: thin line.
 Hypothesized solution as in \er{asympt2} with $v=-0.98$: small broken lines.
 Hypothesized solution with $v=-0.05$: large broken lines. The inset shows the curves near $t=0$.}
 \end{figure}

 \section {Further Extensions}
 To treat non-Markovian processes, the vector potentials have to be functions of the
 $\gamma$ vectors at earlier times, but otherwise, no change in the formalism is needed.

 Non-negativity of the entropy-change follows from the master equations and properties
  of the scattering probabilities ${M_{ba}}$ in \er{diagonal}, as is shown in
 \cite {Kampenbook}.

 Transport processes can be treated simply. Thus, let us consider electronic
 conduction in a solid due to a spatial gradient in the potential (i.e., an electric field)
  or in the ambient temperature. The $\gamma$ vectors are, normally, labeled
 by the reciprocal, $k $-vector  index and are essentially small
 deviations from the square root of the  equilibrium, Fermi-Dirac
 electronic distribution function. Following the standard treatment given in,
 e.g., \cite{Ziman}, the time derivatives of the $\gamma$ vectors (that are
 now real and identical to $\gamma^+$) are proportional to the spatial gradients.
 The vector potentials represent the scattering integral. Then either equation of
 motion in \er{gd} is simply the Boltzmann equation in an inhomogeneous form;
 namely, its left hand side represents the source or the gradient and the right hand
 side contains the desired distribution function under the integral over all
 wave-vectors. The Lagrangian can be used to
 obtain the solution variationally. This variational formulation is, however,
 different from those given in \cite {Kohler,Ziman}. (Of course, different
  variational procedures can lead to the same result.)

 We have noted earlier that the postulated Lagrangian does not contain a
 potential. Adding a potential to the Lagrangian might apparently change the equations
  of motion. It seems to us, however, that under conditions prevailing in stochastic
 processes, this will not happen. The reason for this can be stated in various
  forms and is rooted in the circumstance (already noted above) that in the presence of a random force
 one has no control over the value of the variables, only on its rate of change.
 ("Free terminal endpoint" condition of \cite {Lavenda}, p. 9.)
  In Appendix B we give a formal proof for the following proposition: "When
   {\mbf the following conditions hold: (a) } the potential is a non-negative
 quadratic form in all of its variables (the
 $\gamma$'s), (b) the vector potentials $\bf A $ are all real and positive, (c)
  the initial values of the $\gamma$ vectors are suitably chosen,
 and (d) the variation is performed under conditions of fixed initial values
 of $\gamma$ and $\gd$, then it follows that the action obtained from the variation
  of the velocities only, i.e., with the potential regarded as ignorable,
 is less than the action obtained from the variation of both the variables and
 their derivatives (namely, through the usual Lagrange equations, which are obtained
 under fixed initial and final boundary conditions)."
  The result holds probably under a wider range
 of conditions, since in the proof we have not utilized the requirement imposed
 on the self-correlation of the random forces $f_a$ by the "second" theorem
 of Mori \cite {Mori}. (This requirement ensures, among other things, the
 time-shift invariance of the random process {\mbf which is at the root
 of the Onsager-Machlup theory \cite {O&M, McKean}.} Needless to say,
  that the  result {\mbf obtained in Appendix A} is not in conflict with the validity of
 the Euler-Lagrange equations, since these are obtained under conditions that the
  variables have fixed values at the final time.

 \section{Conclusion}

 The  variational action (or Lagrangian) proposed in \er {action1} for dynamical
 processes has the advantages of being simple, general and flexible. It differs from previously
 employed variational procedures by the factorization ansatz in \er {rho}, by the
  absence of a scalar potential term and the presence of a variable final time upper limit.
  The relation  of the postulated Lagrangian to some basic invariance
 property (like "frame indifference" \cite {JCL}) remains to be explored,
 account being taken of the fact that, for vector potentials that are not all
 equal, the formalism is non-Abelian (namely, the vector potentials cannot be
 transformed away by a single gauge factor)\cite {YM}.
\\
\\
{\bf Acknowledgement}\\

The first named author thanks S. Machlup for a discussion. The application of the
formalism to a (non-linear) evolution was prompted by a referee.

\appendix
 \section {A Tutorial on the Factorized Density Matrix Formalism.}

  Though the factorized density matrix, written in an abstract form as $\rho=
  \gb\cdot\gcrb$, has
 been employed before in \cite {BR} and \cite {SGS}, we shall explain its formalism
  here, following Band's introductory texts to the von Neumann matrix
  method (\cite {Band},\cite{Neumann}).
  Let $\Psi_{\alpha}$ be a possible wave function describing the quantum state
   of the $\alpha$'th system in the ensemble $(\alpha=1,2...N)$. It
   can be expanded in terms of a set of eigenstates $u_n$ as
   \beq
   \Psi_{\alpha} = \sum_n \gamma^{\alpha}_n u_n
   \label {expansion}
   \enq
   As derived in \cite {Band} and other texts, the density matrix arises
   from the ensemble average over all systems in the sense that its
   $nm$ component is
   \beq
   \rho_{nm} = \frac {1}{N} \sum
   _{\alpha} \gamma^{\alpha}_n\gamma^{\alpha +}_m
   \label {rhomatrix1}
   \enq
   The $\gamma^{\alpha}$'s are best viewed as row vectors, distinct for each
   $\alpha$ (or system) and the $\gamma^{\alpha +}$'s as column
   vectors. The $\gamma$- and $\gamma^{+}$- derivatives in the text
 (which implement the variation
   procedure) are with respect to $\gamma^{\alpha}_n$ and $\gamma^{\alpha +}_m$.
     The ensemble averaging, namely the summing over $\alpha$ and the
   subsequent division by $N$, is not explicitly written out in the text, but
    is designated by inserting a dot between $\gamma$ symbols, so that the
     previous matrix element is  written as
   \beq
   \rho_{nm} = \gamma_n\cdot\gamma^{+}_m
   \label {rhomatrix1b}
   \enq
 It is clear that the $\gamma$'s are not vector quantities, but the
 traces over the {\bf dotted} products are proper scalars.

  \section {Proof of the minimal action under one-point boundary condition}

 Assumptions: In the action, \er {action1}, the vector potential $A$ is now assumed to be
  positive (non-negative) and real. We shall further subtract from the action
 (see below) a potential term, in which the potential $V$ depends on the variables
  only, not on their derivatives. This potential is supposed to be monotonic,
  non-decreasing and positive in the relevant range of its variables.

 We start the proof for a single time dependent variable $g$ which replaces
  the earlier complex variable $\gamma$ through
 \beq
 \dot g=\dot g(t)=-i\gd
 \label{g}
 \enq
 The reality of $g$ for all times will be evident. The one-point (initial time)
 boundary conditions fix $g(0)$ and ${\dot g(0)}$, while $g(t)$ at later times
 develops according to its equation of motion. We write the action,
  including the potential, in the single variable $g$ as
 \beq
  S(T) = \int_0 ^{T} dt [({\dot g}- A(t))^2 -2 V(g(t))]
 \label {action2}
 \enq
 The boundary conditions fix the value of $g(0)>0$ and of $\dot g(0)$. We next
  minimize the above action in two ways and subsequently compare the
 resulting actions. The first is the usual Lagrange equation way in the presence
 of a potential $V$ and the quantities arising from this method will be denoted by
  the superscript $V$. The second method pretends that there is no potential
 and the corresponding quantities will take the superscript $0$. It is
 the second method that was used in the  text.
 \beq
 \ddot g^V (t) = \dot A (t) - V'(g(t))
 \label{gV}
 \enq
 where the prime represents the derivative with respect to the argument  $g$.
 \beq
 \dot g^0 (t) =  A(t)
 \label {g0}
 \enq
 The latter equation imposes the following initial condition for the velocities:
 \ber
 \dot g^0 (0)& = & A(0)
 \nonumber\\
 & = & \dot g^V (0) \label{gdot0}
 \enr Integrating \er {gV}
once, we obtain
\beq
\dot g^V (t) =  A(t) - P(t) \qquad
P(t) \equiv \int_0^t V'(g(t))dt
\label {gdotV}
\enq
where $P(t)$ is zero at $t=0$ and is for positive times non-negative,
since it is, by \er {gV}, the time integral of a positive quantity.
Subtracting $ \dot g^0 $ shown in \er {g0} from the last equation and integrating,
it is clear that $g^V$ never exceeds (algebraically) $g^0$. Calculating the
actions obtained in the two methods and subtracting we find:
\ber
S^V (T) &=& \int_0 ^{T} dt [({\dot g^V}- A(t))^2 - 2 V(g^V (t))] = \int_0 ^{T} dt [P(t)^2 - 2 V(g^V (t))]
\nonumber \\
S^0 (T) &=& \int_0 ^{T} dt [({\dot g^0}- A(t))^2- 2 V(g^0 (t))] =\int_0 ^{T} dt [- 2 V(g^0 (t))]
\nonumber \\
&\Rightarrow& S^V (T) -S^0 (T) = \int_0 ^{T} dt [P(t)^2 +2 \bigl( V(g^0 (t))-V(g^V (t))\bigr)]
\label {Sdiff}
\enr
In the integrand the squared term is necessarily positive (non-negative)
and so is the term containing the difference of potentials since the 0-argument
is larger than the V-argument and the potential is monotonic by supposition.
Though obtained under restricted conditions, the result shows clearly that
 the two-point boundary conditions are necessary requisites for the validity
of the Lagrange-Euler equations of motion.
Generalization to several (real) variables $g_1, g_2, ..., g_N $ is immediate, when
the potential is a positive quadratic form in these variables, since
this can be diagonalized (with positive eigenvalues) simultaneously with the
kinetic energy term. However, the initial point variables need to be chosen
carefully in this case.

Finally, we have not proven that the action using equations of motion of the text
is minimal, but only that is lower than that obtained with the (for this case,
 inappropriate) use of the Lagrange equations. Furthermore, it is not evident
 that the solutions obtained in this Appendix satisfy conditions required from
density matrices or probabilities (e.g., normalizations).

\end{document}